\documentclass[slac_one]{revtex4}
\usepackage{graphicx}
\usepackage{fancyhdr}
\usepackage{floatflt}
\usepackage{epsfig}
\begin{document}

\title{CKM global fit and constraints on New Physics in the $B$ meson mixing  } 

%

\author{O. Deschamps [CKMFitter group]}
\affiliation{Laboratoire de Physique Corpusculaire, IN2P3/CNRS, Clermont-Ferrand, France}

\begin{abstract}
An update profile of the CKM matrix is given, providing numerical and graphical constraints on the CKM parameters in the Standard Model. Constraints on additional parameters accounting for possible new physics contribution in a model-independent analysis are also reported
with emphasis on the leptonic decay of the $B_d$ and on the $B_s$ mixing phase.
\end{abstract}
\maketitle

\thispagestyle{fancy}

\section{INTRODUCTION} 
Within the Standard Model (SM), the quark flavor mixing is described by the Cabibbo-Kobayashi-Maskawa (CKM) matrix~\cite{CKM}. The 3x3 unitary CKM matrix can be fully parametrized by four independent parameters among which a single non-vanishing complex phase accounts for the violation of the CP symmetry. \\
Inspired from the one proposed by Wolfenstein~\cite{Wolfenstein} the following parametrization, phase-convention independent and unitary-exact to all orders in $\lambda$~\cite{thePapII}, 
 is used throughout this document:
\begin{eqnarray}
\lambda = \frac{|V_{us}|}{\sqrt{|V_{ud}|^2+|V_{us}|^2}},~    A\lambda^2=\frac{|V_{cb}|}{\sqrt{|V_{ud}|^2+|V_{us}|^2}},~  
\bar\rho + i\bar\eta= - \frac{V_{ud}V_{ub}^*}{V_{cd}V_{cb}^*}.	\nonumber
\end{eqnarray}
While the parameters $\lambda$ and $A$ are  accurately determined  through the  measurement of $|V_{ud}|$ from the super-allowed nuclear transitions, $|V_{us}|$ from the semileptonic kaon decay  and $|V_{cb}|$  from the semileptonic $B$ decay with charm, the $(\bar\rho + i\bar\eta)$ complex parameter,  being the apex coordinates of the unitarity triangle (UT) related to the first and third quark families, is less constrained. The metrology of the UT apex via the determination of its sides and angles allows to measure the size of the CP violation and to validate the overall consistency of the KM scenario within the Standard Model. Any inconsistency would suggest contributions from physics beyond the Standard Model.

\section{THE CKM FIT INPUTS}
s
The CKM global fit is performed within a frequentist statistical approach including a specific  treatment to handle the theoretical uncertainties on some of the inputs (RFit~\cite{thePapII}). The inputs used for the fit are the most recent available value of the  observables from the $K$ and $B$ meson  physics for which the theoretical conversion into CKM parameters is under control.
The most relevant input observables of the global CKM fit, with their value and reference, are listed in the table~\ref{tab:inputs}. The observable values that have been updated for this analysis with respect to the previous similar analysis~\cite{CKMMoriond08} are indicated.

\begin{table}[bh]\begin{footnotesize}
\begin{tabular}{|c|l|lc|c||c|c|}
\hline        
           input   & source  & value & (ref)  & update & UT constraint & associated theory  [parameters] \\
\hline
\hline
$|V_{ud}|$    &  Nuclear decays &  $0.97418\pm0.00026$&(\cite{Towner}) & & $|V_{ud}|$  & [$|V_{ud}|$ form factor]   \\
$|V_{us}|$    &  SL kaon decays    &  $0.2246\pm0.0012$&(\cite{Flavia}) & &$|V_{us}|$   & [$|V_{us}|$ form factor]   \\
$|V_{cb}|$ &   SL charmed $B$ decays & $(40.60\pm0.35\pm0.58)10^{-3}$&(\cite{HFAG}) & &$|V_{cb}|$ & [$|V_{cb}|$ form factor and/or OPE]  \\
$|V_{ub}|$   &   SL charmless $B$ decays &$(3.76\pm0.10\pm0.47)10^{-3}$&(\cite{HFAG})& &$|V_{ub}|$ & [$|V_{ub}|$ form factor and/or OPE]  \\
BR$(B^+\to\tau^+\nu)$  & Leptonic B decays & $ (1.73\pm0.35)10^{-4}$&(\cite{TauNu}) & *&  $|V_{ub}|$ & leptonic  amplitude  [$f_{B_s}$,  $f_{B_s}/f_{B_d}$] \\	
$\Delta m_s$  & $B_s\bar B_s$ mixing & $(17.77\pm0.12)\textrm{ps}^{-1}$&(\cite{Dms}) & & $|V_{ts}V_{tb}^*|$ & $\Delta B$=2 amplitude  [$B_s$, $f_{B_s}$, $\bar m_t$ , $\eta_{B}$] \\ 
$\Delta m_d$ &  $B_d\bar B_d$ mixing & $(0.507\pm0.005)\textrm{ps}^{-1}$&(\cite{HFAG}) & &$|V_{td}V_{tb}^*|$ & $\Delta B$=2 amplitude  [$\hat B_{B_s}/\hat B_{B_d}$ $f_{B_s}/f_{B_d}$, $\eta_{B}$]  \\
\hline
\hline
$|\epsilon_K|$ & $K\bar K$ mixing &  $(2.229\pm0.010)$&(\cite{PDG}) & * &  $f(\bar\rho,\bar\eta$) & $\Delta S$=2 amplitude  [$B_K$, $\eta_{B}$ , $\eta_{ct}$, $\eta_{tt}$] \\
sin$(2\beta)$ & Charmonium $B$ decays  &  ($0.672\pm0.024$) &(\cite{HFAG}) & * & $\beta$ & -    \\
BR \& $A_{CP}$ & $B\to\pi\pi,\rho\rho,\rho\pi$  decays    &  B-factories average &(\cite{angle})   & * &$\alpha$& SU(2)    \\
BR \& $A_{CP}$ &  $B\to DK$  decays  &  B-factories average&(\cite{angle})  & * & $\gamma$ & GLW/ADS/GGSZ   \\
 \hline
\end{tabular}
\caption{Most relevant inputs for the global CKM fit. The left part addresses the observable inputs with their value and reference. The right part addresses the constraint on UT parameters derived from the observable. The upper part of the table contains CP-conserving observables  while the CP-violating are listed in the lower part.
\label{tab:inputs}}
\end{footnotesize}
\end{table}

\begin{floatingtable}[r]{
\begin{tabular}{|c|c|c|}
\hline
 observables &  values used for this analysis~\cite{Tantalo,Nierste}& possible update\cite{Lubicz}\\
\hline
$f_{B_s}$            &   $(268\pm 17\pm 20)$ MeV& $(245\pm 25)$ MeV\\
$f_{B_s}/f_{B_d}$    &   $1.20\pm 0.02\pm 0.05$&$1.21\pm 0.04$ \\
$\hat B_{B_s}$                &   $1.29\pm 0.05\pm 0.08$& $1.22\pm 0.12$ \\
$\hat B_{B_s}/\hat B_{B_d}$            &   $1.00\pm 0.02$&$1.00\pm 0.03$\\
$\hat B_K$                &   $0.78\pm 0.02\pm 0.09$&$0.75\pm 0.07$ \\
\hline
\end{tabular}}
\caption{Phenomenological quantities obtained from the LQCD calculation.}\label{tab:lattice}
\end{floatingtable}

The right side of the table addresses the associated theory, possibly including phenomenological para\-me\-ters, needed to derive  a constraint on the UT parameter(s) from the input observable.
The value coming from the lattice QCD calculation (LQCD) for some of the phenomenological quantities  used in this analysis  are listed in the second column of the table~\ref{tab:lattice}. For comparison purpose, some alternative and more recent values proposed in the reference~\cite{Lubicz} are quoted in the last column (note that the statistical and theoretical uncertainties have been assumed quadratically summable by the authors).\\
A more detailed review of the CKM global fit inputs can be found in~\cite{CKMfitter}.

\section{THE SM CKM FIT RESULTS}

\begin{figure}[hb]
\begin{flushleft}
\includegraphics[width=85mm,height=83mm]{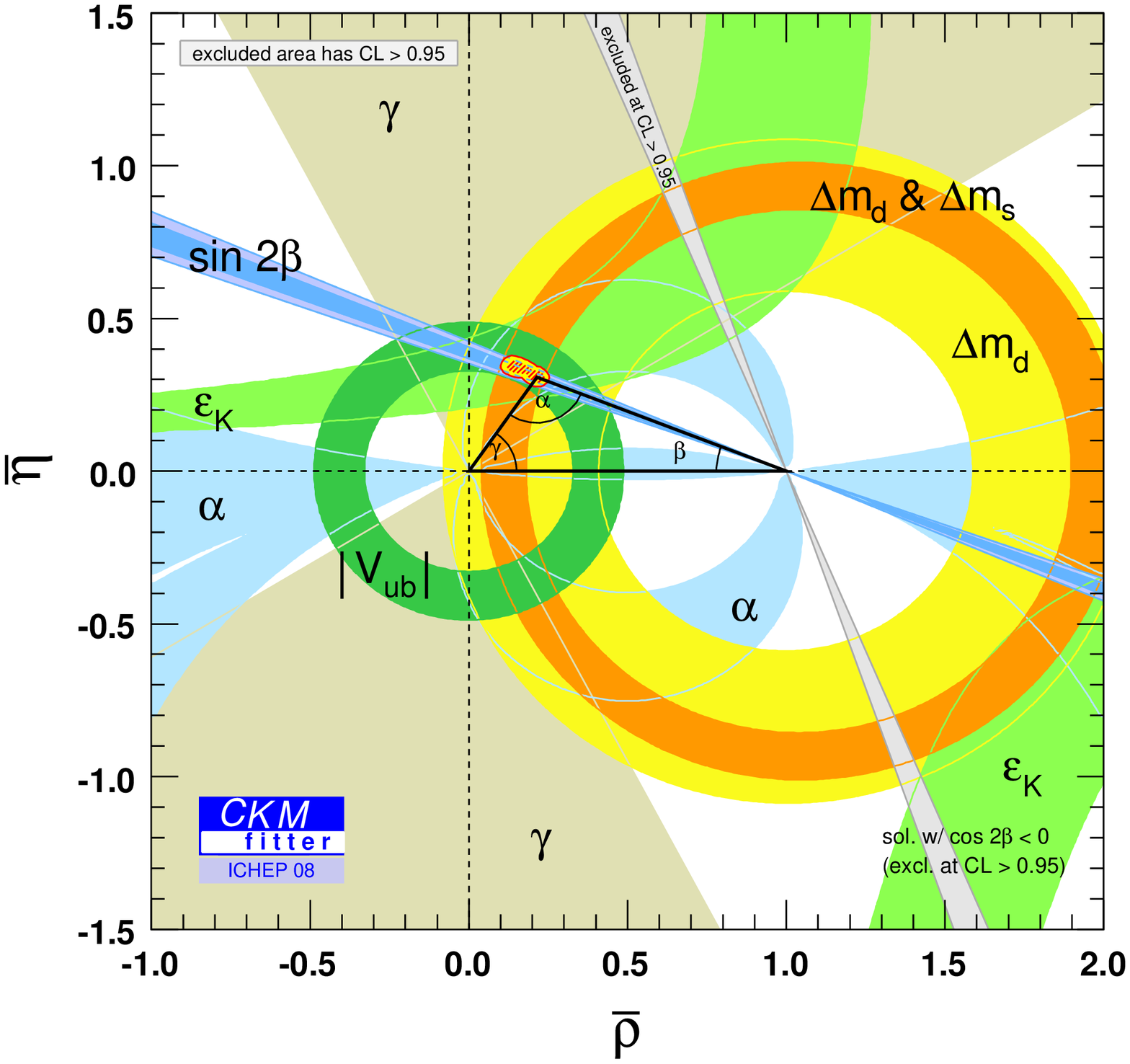}\begin{minipage}[b]{75mm} 
\begin{flushright}  
\includegraphics[width=85mm,height=41mm]{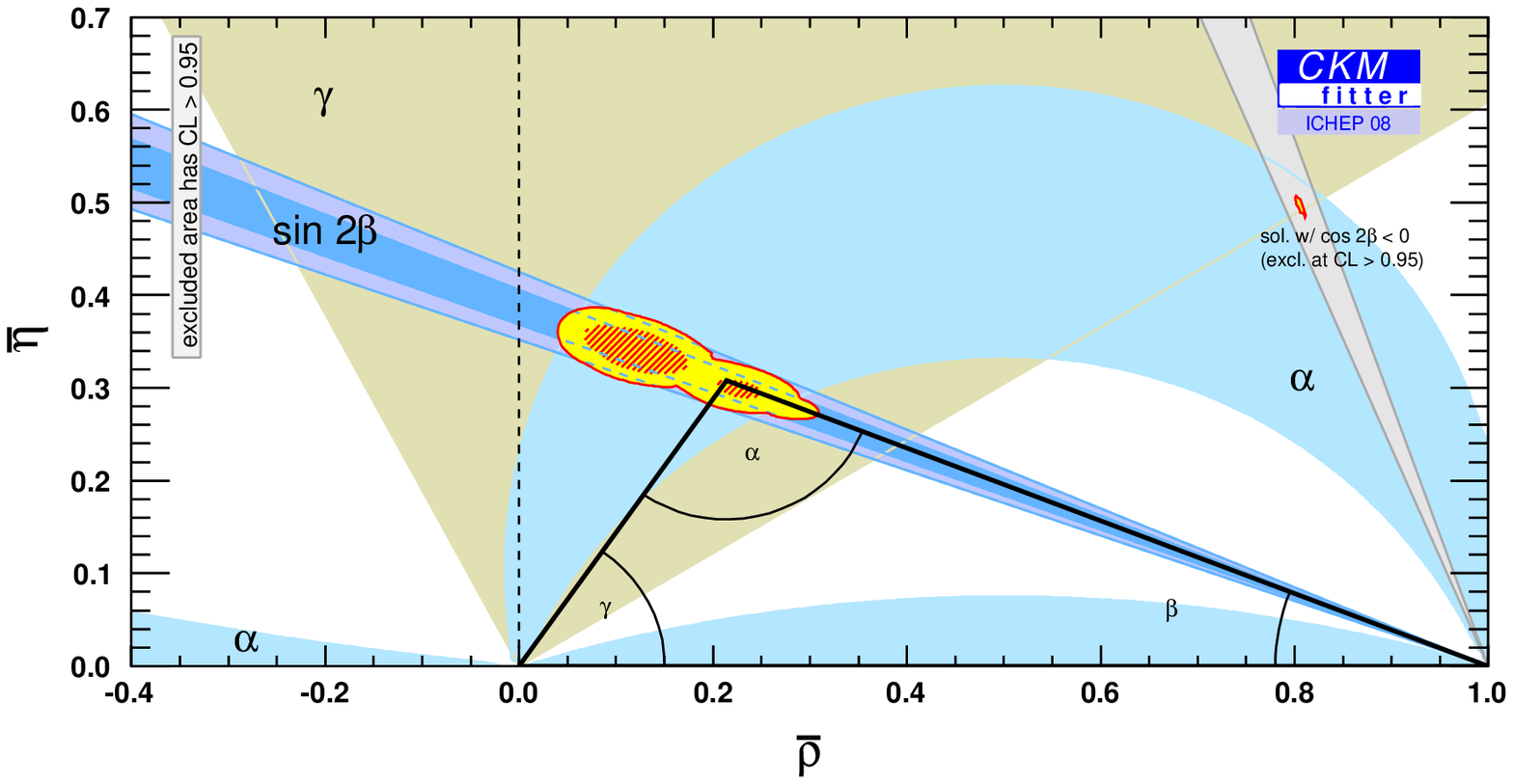}
\includegraphics[width=85mm,height=41mm]{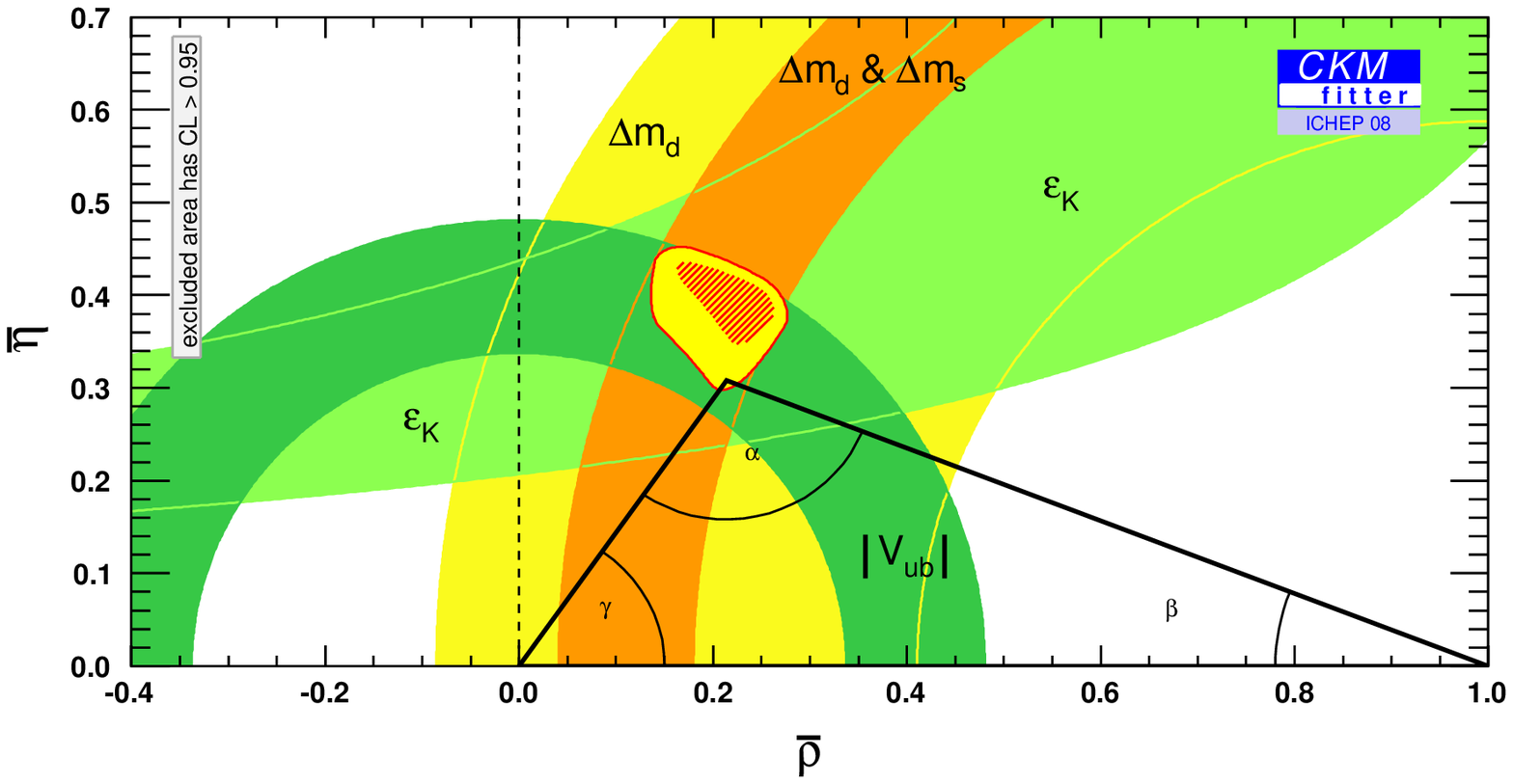}
\end{flushright}  
\end{minipage}
\end{flushleft}
\caption{Left figure: individual and global constraints in the ($\bar\rho$,$\bar\eta$) plane from the global CKM fit. The  hashed region of the global combination corresponds to 68\% CL. The constraints from the experimentally dominated observables and observables inducing larger theoretical uncertainties are shown on the right-top and the right-bottom figure, respectively. 
\label{fig:global}}
\end{figure}

The left part of the figure~\ref{fig:global} displays the result of the global fit together with the 95\% CL contours of the individual constraints.

\vspace{.5cm}

 A satisfactory agreement is observed from the various individual contributions at the 2$\sigma$ level, establishing the KM mechanism as the dominant source of the CP violation in the $B$ meson system.\\
 As shown on the right side of the figure~\ref{fig:global}, a slight tension is however revealed when comparing the global constraint coming from the observables dominated by the experimental measurements  (UT angles) with the constraint derived from the observables inducing larger theoretical uncertainties ($|V_{ub}|$,  BR($B^+\to\tau^+\nu$), $|\epsilon_K|$, $\Delta m_{d}$/$\Delta m_{s}$). This classification of the observables is for illustration purpose and should not be used to draw any conclusion about the possible  origin of the tension: a similar discrepancy is also visible when comparing the constraint  from the tree-level processes  with the mixing loop-induced  or the constraint from the CP-conserving observables versus the CP-violating ones.
 
\begin{figure}[h]
\includegraphics[width=55mm,height=69mm]{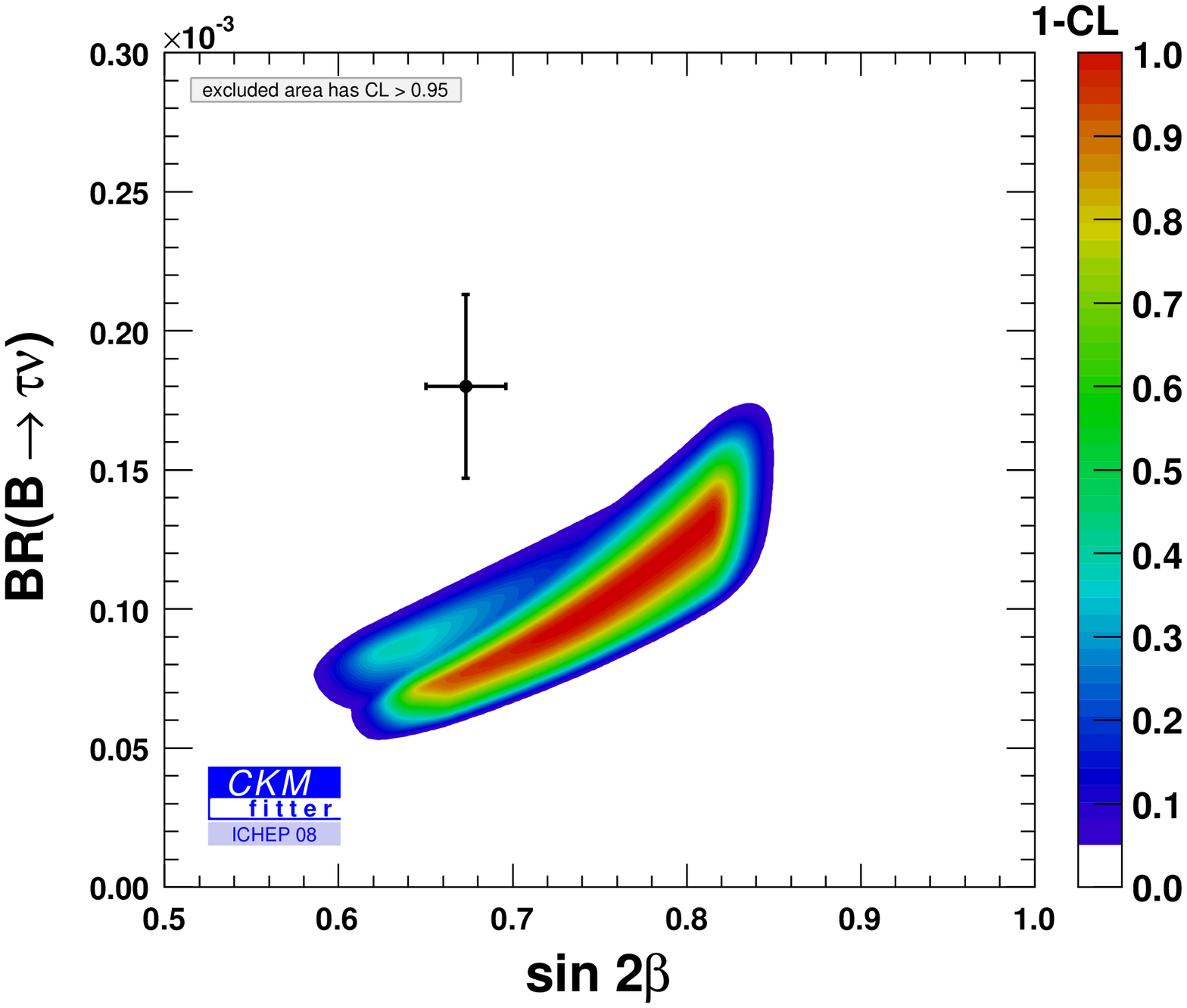}
\includegraphics[width=55mm,height=69mm]{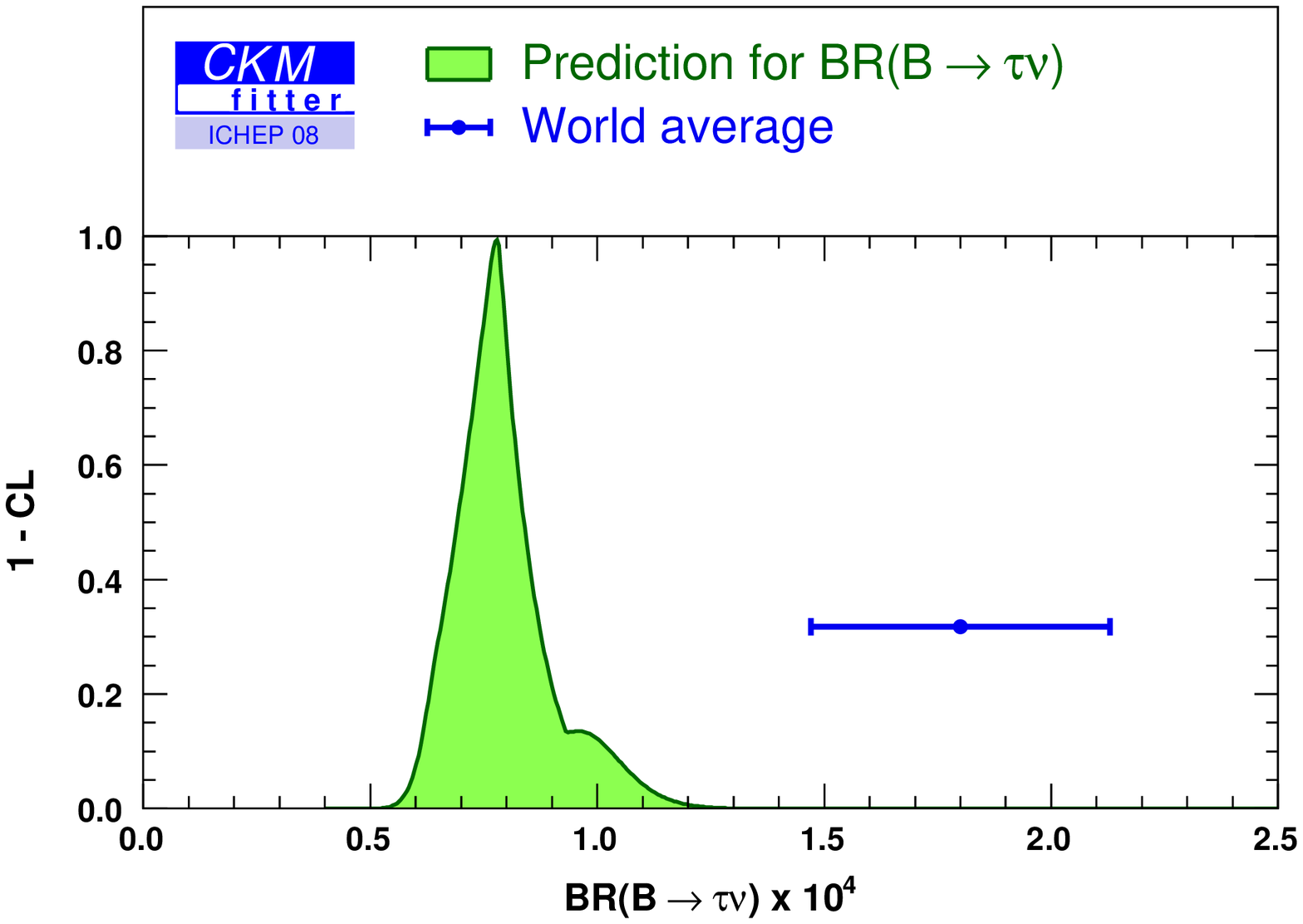}
\includegraphics[width=55mm,height=69mm]{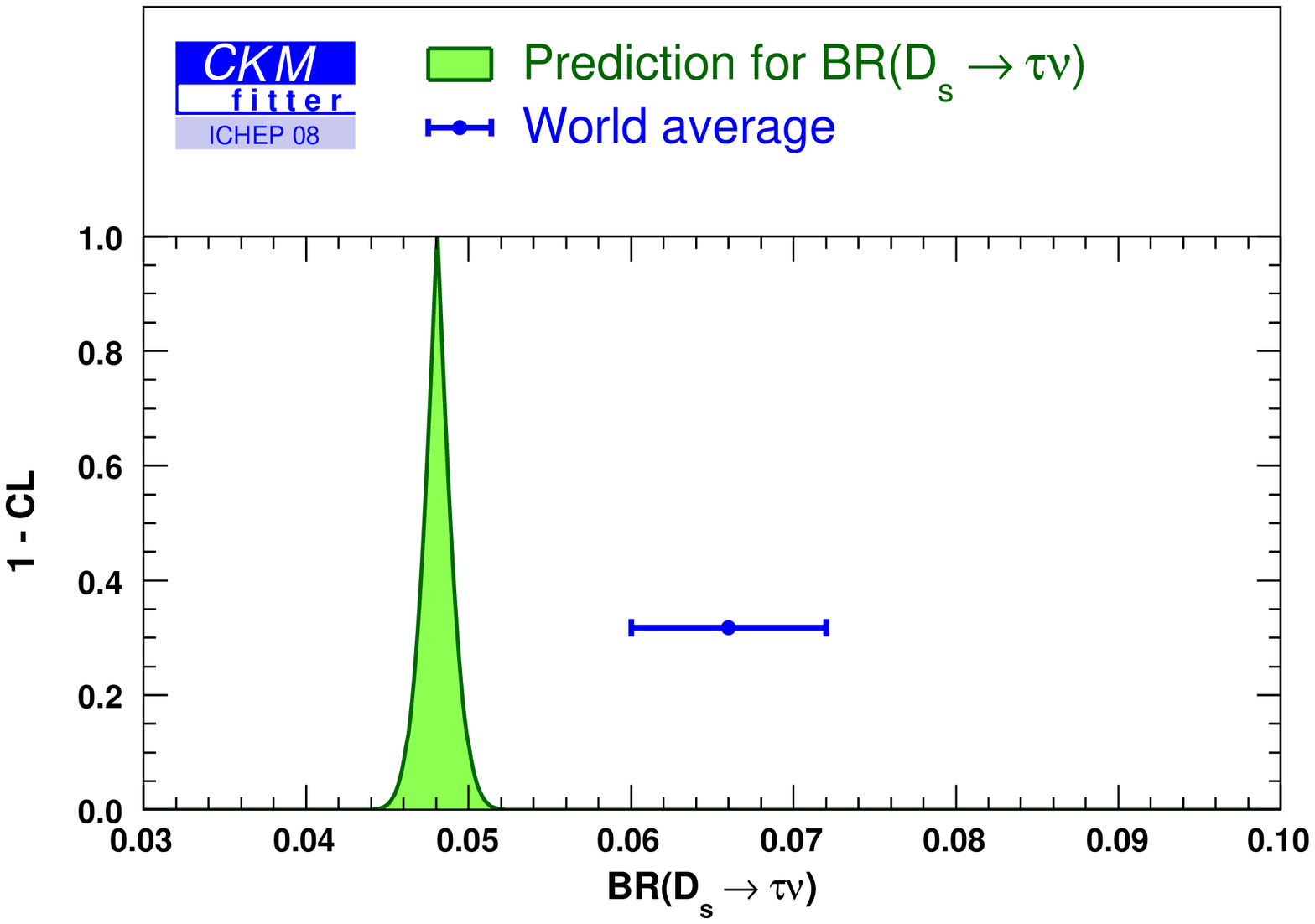}
\caption{Left: Confidence Level contour in the 2D (sin$(2\beta)$,$BR(B^+\to\tau^+\nu)$) plane for the fit prediction. The cross indicates the 68\% CL interval for the direct experimental measurement. 
Middle: the corresponding 1D projection for BR$(B^+\to\tau^+\nu)$ compared to the direct measurement. For comparison purpose, the prediction for the branching ratio of the $D_s\to\tau^+\nu$ leptonic decay is displayed on the right plot (the decay constant value $f_{D_s}=241 \pm 3$ MeV has been used~\cite{Follana}).\label{Fig:tension} }
\end{figure}
 
 As illustrated on the left plot of the figure~\ref{Fig:tension}, the bulk of the tension is located in the  correlations between the CP-conserving, UT-side related, theory-dependent observable, BR$(B^+\to\tau^+\nu)$  (central value of which slightly increases with the summer 2008 update~\cite{TauNu}) and the CP-violating, UT-angle related, theory-free observable, sin$(2\beta)$ (central value of which slightly decreases~\cite{HFAG}).\\
Quantitatively, the minimal $\chi^2$ of the global fit  decreases by 2.9~$\sigma$ (2.6~$\sigma$) when removing  BR$(B^+\to\tau^+\nu)$ (sin$(2\beta)$) from the list of the fit inputs. Obviously, the disagreement is not large enough to exclude the possibility of a statistical origin of the tension.
However, it is worth mentioning  that a similar discrepant pattern  is also observed for the $D_s$ leptonic decays (see for instance ~\cite{Kronfeld,Gamiz}), suggesting a common origin. 
As an illustration, the expected and measured branching fraction of the leptonic $B^+$ and $D_s$ decays to $\tau^+\nu$ final state are displayed on the right side of the figure~\ref{Fig:tension}. 
\begin{figure}[hbt]
\includegraphics[width=60mm,height=60mm]{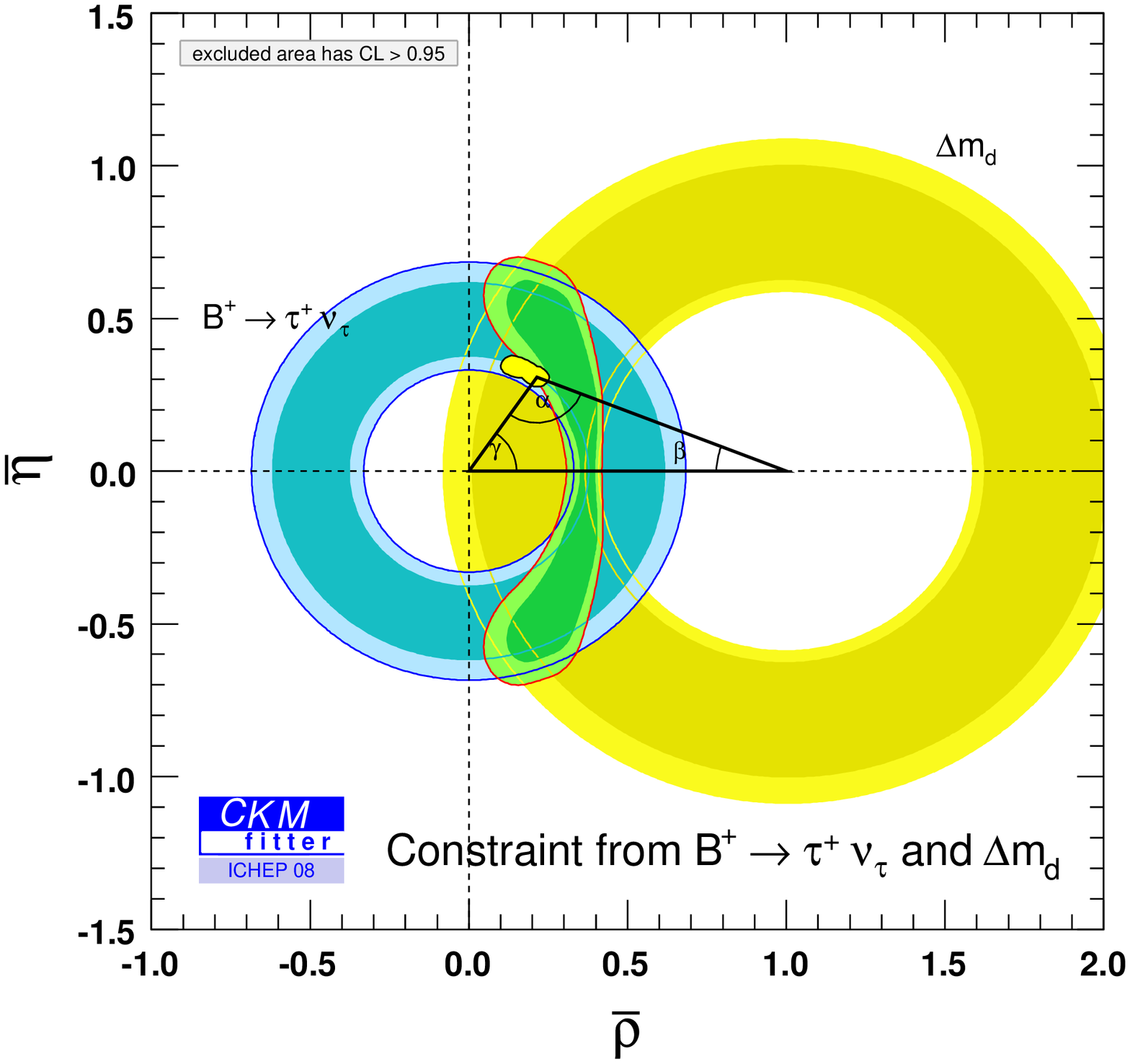}\includegraphics[width=60mm,height=66mm]{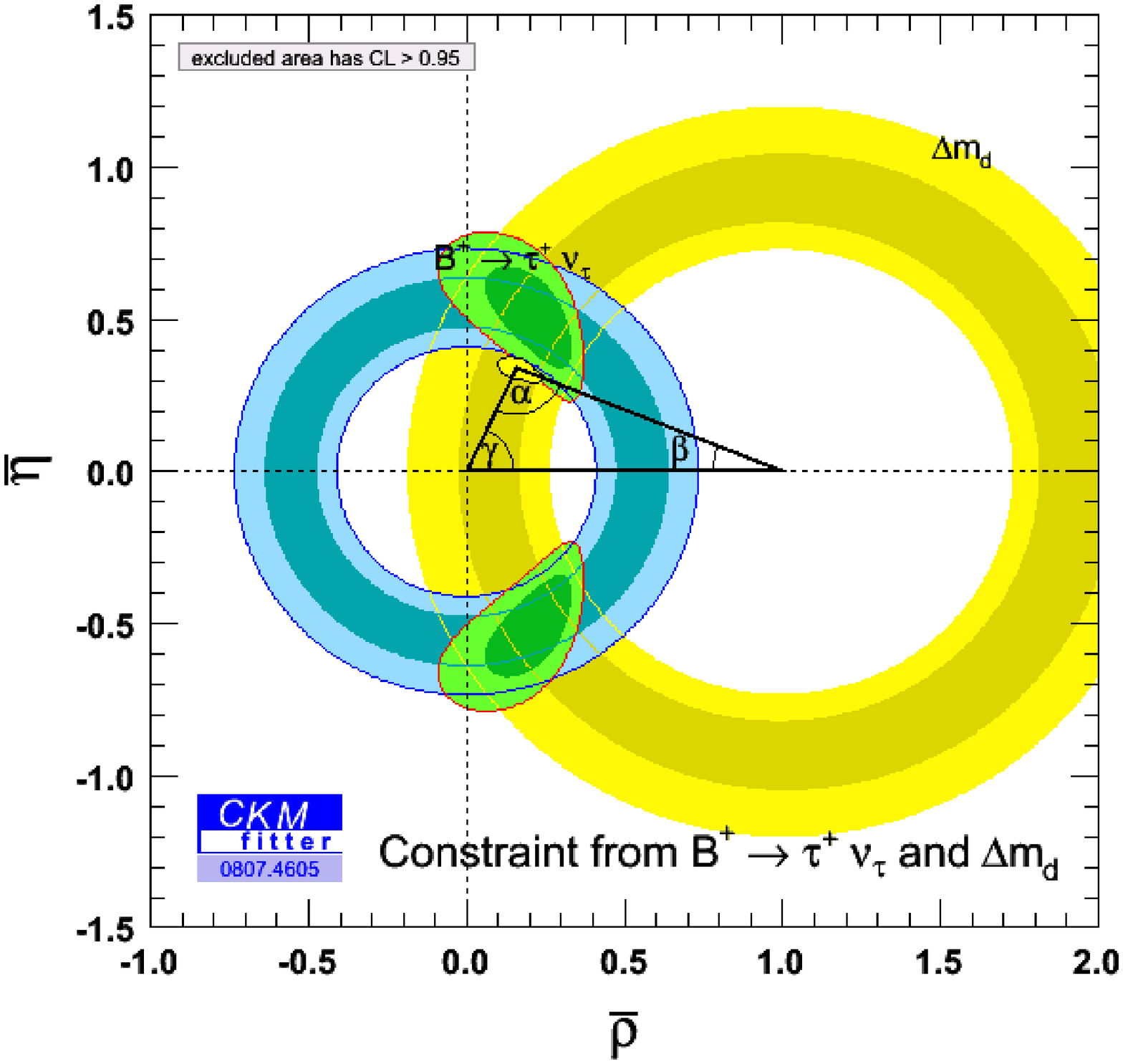}\includegraphics[width=60mm,height=62mm]{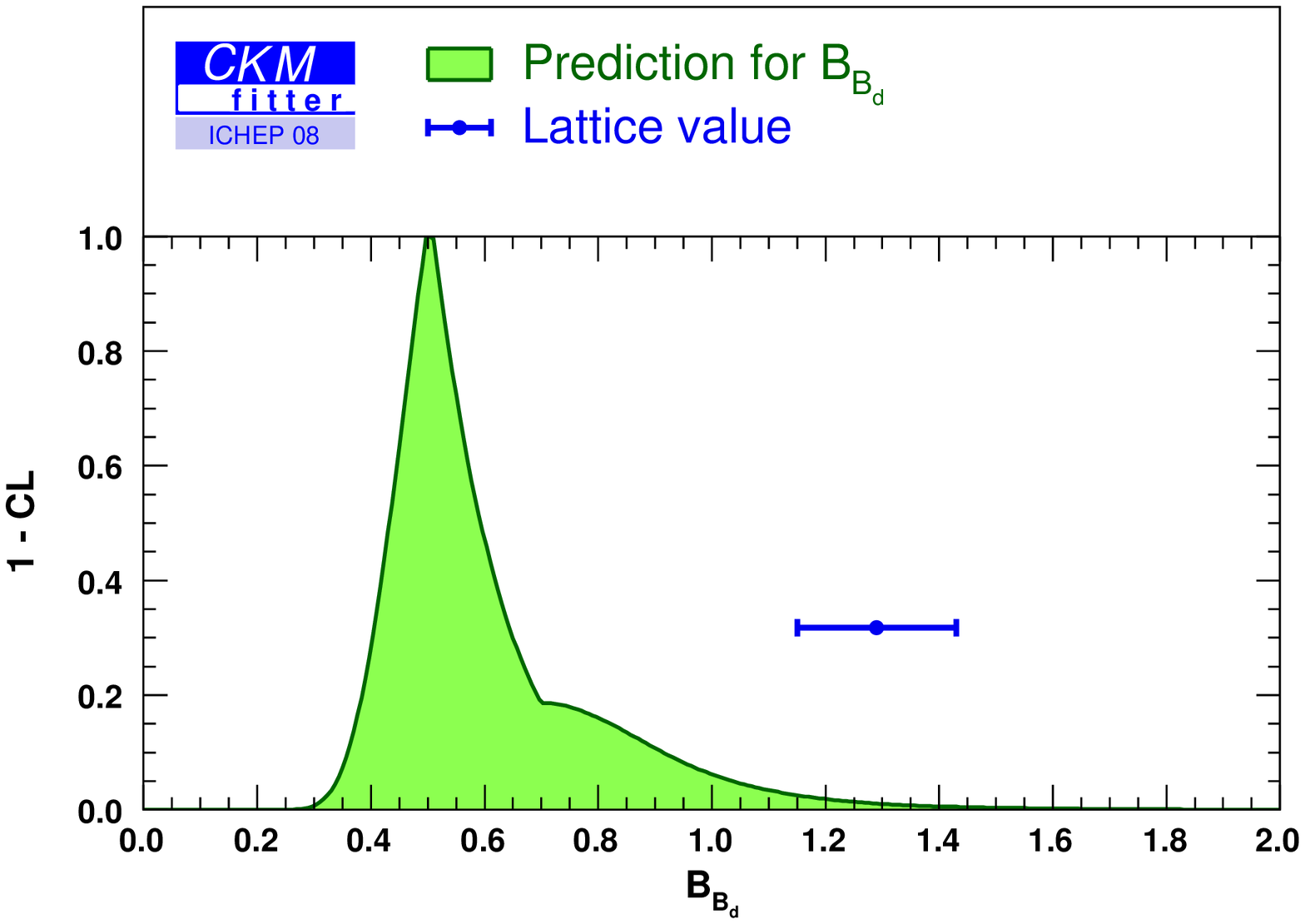}
\caption{Constraint in the ($\bar\rho$,$\bar\eta$) plane from the BR$(B^+\to\tau^+\nu)$ and $\Delta m_{d}$ observables only using the LQCD parameters value quoted in the 2nd column of the table~\ref{tab:lattice} (left) and using the alternative values quoted in the third column (middle). 
In the latter case, both the theoretical and statistical uncertainties are assumed to be Gaussianly distributed, resulting in a more aggressive impact of the LQCD parameters.
On both plots the global constraint from the whole set of observables is indicated by the shaded area around the apex of the unitary triangle. Right: theory-free prediction for the bag parameter $\hat B_{B_d}$ (see text) compared to the LQCD calculation quoted in the second column of the table~\ref{tab:lattice}.
\label{fig:Bag} }
\end{figure}
The possibility of a generic problem with the lattice prediction for the theoretical parameters related to the $B$ meson has  been investigated. 
 Although the tree-level expression for the amplitude of the the $B^+\to\tau^+\nu$ annihilation decay is directly proportional to the decay constant $f_{B_d}$, it can be shown that an under-estimation of the lattice prediction for the parameters product $f_{B_s}\times({f_{B_d}}/{f_{B_s}})$ as used in this analysis,  can not by itself explain the observed tension.
Indeed, the BR$(B^+\to\tau^+\nu)$ together with the $\Delta m_d$ observables provide a $f_B$-independent constraint in the ($\bar\rho ,\bar\eta$) plane in which a clear tension remains as illustrated on the   figure~\ref{fig:Bag} for the two set of the LQCD parameters values quoted in the table~\ref{tab:lattice}. \\
More precisely, the $f_B$-independent ratio: 
\begin{eqnarray}
\frac{\textrm{BR}(B^+\to\tau^+\nu)}{\Delta m_d} = \frac{3\pi}{4}\frac{m_\tau^2 }{m_W^2 S(x_t)}\left(1-\frac{m_\tau^2}{m^2_{B^+}}\right) ^2 \tau_{B^+} \left(\frac{\sin\beta}{\sin\gamma} \right) ^2\frac{1}{|V_{ud}|^2}\frac{1}{\hat B_{B_d}}\nonumber
\end{eqnarray}
allows for a theory parameter-free prediction of the bag parameter, $\hat B_{B_d}$, from the experimental measurement of  $\beta$, $\gamma$ (or $\alpha=\pi-\beta-\gamma$), $|V_{ud}|$, BR$(B^+\to\tau^+\nu)$ and $\Delta m_d$.  The resulting prediction is  2.4 $\sigma$ away from the lattice calculation~\cite{Tantalo,Nierste} as shown in the right plot of the figure~\ref{fig:Bag}.
This deviation is essentially dominated by the experimental uncertainties on the $B^+\to\tau^+\nu$ branching ratio and on the UT angles $\alpha$ and $\gamma$.\\
A lattice origin of the tension would thus involve the bag parameter $B_{B_d}$ that 
controls the correlations between BR$(B^+\to\tau^+\nu)$ and the angle $\beta$, as explicitely shown in the above formula. Further investigations are needed to check whether the possible correlations between the theoretical errors affecting the lattice parameters, generally not provided by the lattician community and therefore neglected in this analysis, could account for the observed tension.\\
Eventually, the hypothesis of new contributions beyond the SM is discussed in the next section.

\section{CONSTRAINTS ON NEW PHYSICS IN THE $B_{d,s}$ MESONS MIXING}
\subsection{Fit procedure}
New Physics (NP) is expected to affect the amplitude of the neutral $B$  mesons mixing in many  scenarii (see for instance ~\cite{Fleisher}). Assuming NP to contribute mostly to the short-distance part of the $\Delta F=2$ processes, a model-independent parametrization has been proposed~\cite{Lenz,Grossman}:
\begin{eqnarray}
\langle B_{q}|M_{12}^{\textrm{\tiny SM+NP}}|\bar B_{q}\rangle = \Delta_q^{\textrm{\tiny NP}} \langle B_{q}|M_{12}^{\textrm{\tiny SM}}|\bar B_{q}\rangle \nonumber
\end{eqnarray}
where the label $q$ stands for the $d$ or $s$ flavor of the neutral $B$ meson and the complex parameter $\Delta^{\textrm{\tiny NP}}_q = |\Delta^{\textrm{\tiny NP}}_q|e^{i\Phi^{\textrm{\tiny NP}}_q}$ accounts for the NP contribution.
Assuming in addition that the tree-level mediated decays proceeding through a Four Flavor Change get only SM contributions (SM4FC hypothesis~\cite{SM4FC,CKMfitter}), the observables $|V_{ij}|$, $\gamma$ and  $\gamma(\alpha)=\pi-\beta_{c\bar c}-\alpha$~\footnote{where $\beta_{c\bar c}$ means the $\beta$ UT angle extracted from the analysis of the charmonium $B_d$ meson decays.}
are not affected by the NP contribution and can be used in a (SM+NP) global fit to fix the SM parameters. The oscillations parameters, the weak phases, the semileptonic asymmetries and the $B$ meson lifetime-differences are affected by the phase and/or the amplitude of the NP contribution as quoted in the table~\ref{tab:NP} and allow to constrain the NP deviation to SM parametrized with $\Delta_q^{\textrm{\tiny NP}}$.

\begin{table}[h]
\begin{tabular}{|c|c|}
\hline
   parameter &  prediction in the presence of NP\\
   \hline
$\Delta m_q$                                    & $|\Delta^{\textrm{\tiny NP}}_q|\times\Delta m_q^{\textrm{\tiny SM}}$ \\
$2 \beta$                                       & $2 \beta^{\textrm{\tiny SM}}+\Phi_d^{\textrm{\tiny NP}}$\\
$2 \beta_s$                                     & $2 \beta^{\textrm{\tiny SM}}_s-\Phi_s^{\textrm{\tiny NP}}$\\
$2 \alpha$                                      & $2 (\pi-\beta^{\textrm{\tiny SM}}-\gamma)-\Phi_d^{\textrm{\tiny NP}}$\\
$\Phi_{12,q} = \textrm{Arg}[-\frac{M_{12,q}}{\Gamma_{12,q}}]$ & $\Phi_{12,q}^{\textrm{\tiny SM}} + \Phi_q^{\textrm{\tiny NP}}$\\
$A_{SL}^q$                                      & $\frac{\Gamma_{12,q}}{M_{12,q}^{\textrm{\tiny SM}}}\times \frac{\sin(\Phi_{12,q}^{\textrm{\tiny  SM}} + \Phi_q^{\textrm{\tiny NP}})}{|\Delta^{\textrm{\tiny NP}}_q|}$\\
$\Delta\Gamma_q$                                      & $2 |\Gamma_{12,q}| \times \cos(\Phi_{12,q}^{\textrm{\tiny SM}} + \Phi_q^{\textrm{\tiny NP}})$\\
\hline
\end{tabular}
\caption{Theoretical prediction for the $B$ physics observables in the presence of NP in mixing. Note the opposite sign convention for the $B_d$ and $B_s$ mixing phase.\vspace{0.2cm}}\label{tab:NP}
\end{table}

\begin{figure*}[hbt]
\centering
\includegraphics[width=75mm,height=80mm]{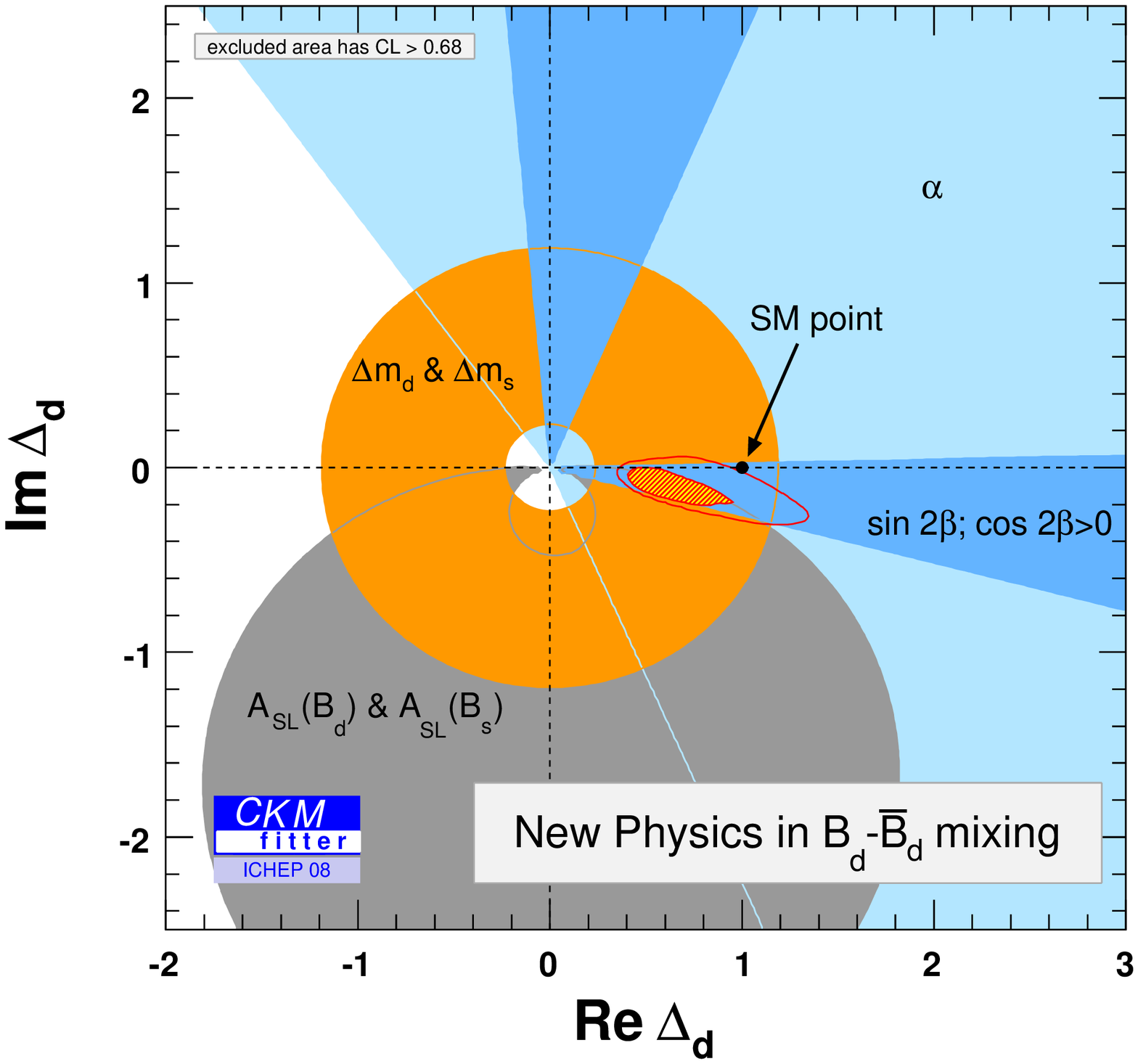}
\includegraphics[width=75mm,height=80mm]{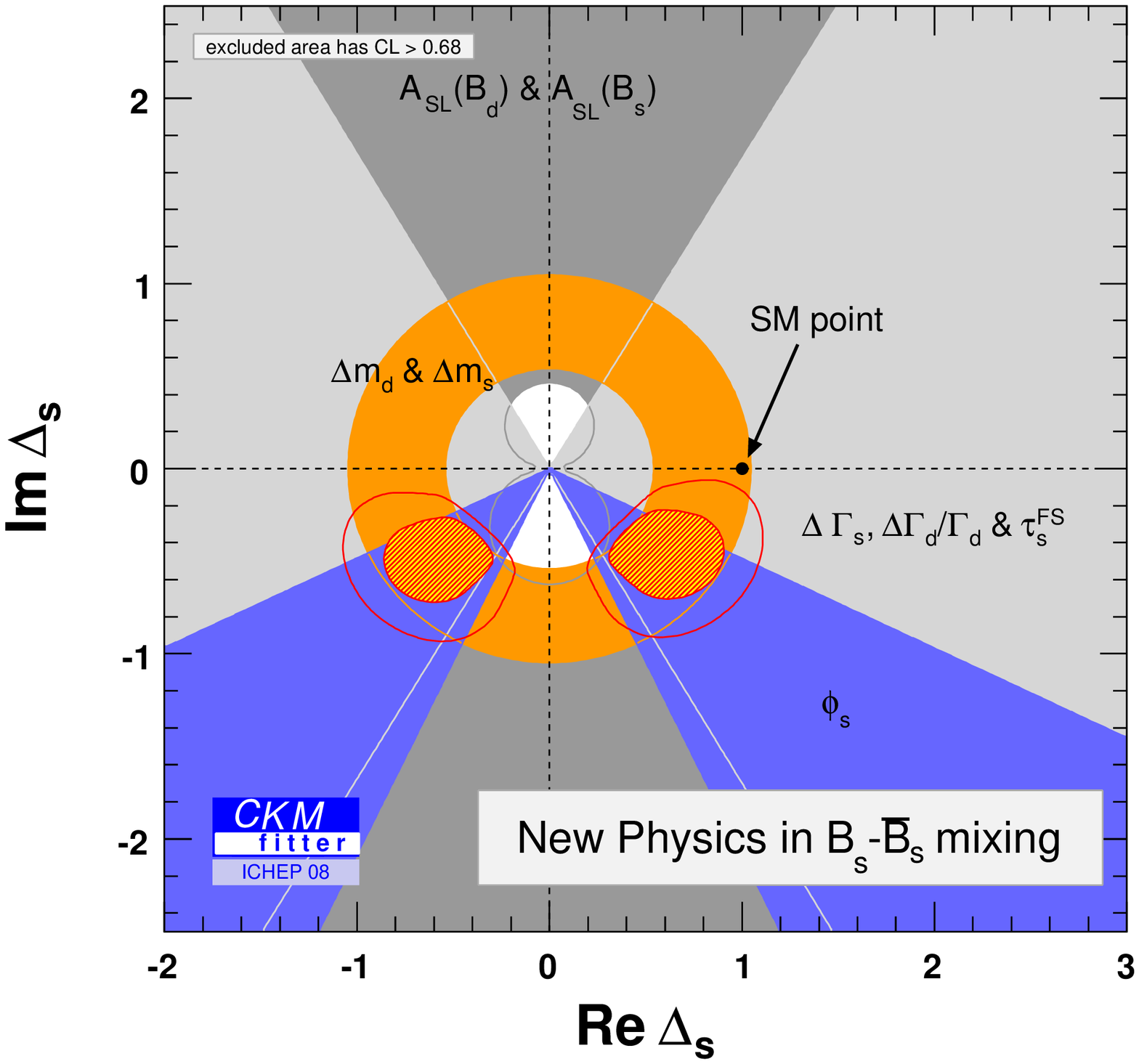}
\caption{Individual and global constraints  in the  (Re$(\Delta_q^{\textrm{\tiny NP}})$,Im$(\Delta_q^{\textrm{\tiny NP}})$) plane for the  (SM+NP) fit. Due to the large uncertainties, only the 68\% CL are shown for the individual constraints. Both 68\% and 95\% contours are shown for the global constraint.} \label{fig:NP}
\end{figure*}

The resulting constraints in the (Re$(\Delta_q^{\textrm{\tiny NP}})$,Im$(\Delta_q^{\textrm{\tiny NP}})$) plane are summarized in the left and right side of the figure~\ref{fig:NP} for the $B_d$ and $B_s$ case, respectively. For a more complete review of the constraints from the global fit of NP in the B meson mixing see~\cite{Capri}.

\subsection{The $B_d$ case}
For the $B_d$ meson case, the dominant constraints come from the mixing phase, $\beta$, and the oscillation frequency $\Delta m_d$. Both reasonably agree with their SM prediction. A 2.1~$\sigma$ deviation is obtained for the 2-dimensional SM hypothesis $\Delta_d^{\textrm{\tiny NP}}=1$. The 1-dimension hypothesis based on the phase only, $\Phi^{\textrm{\tiny NP}}_d$=0,  corresponding the phase expectation in the Standard Model as well as in the  Minimal Flavour Violation scenarii, results in a 1.5~$\sigma$ deviation. This slight disagreement with the SM hypothesis is essentially the translation of the tension in the SM global fit discussed in the previous section (obviously the discrepancy in the $D_s$ leptonic decays  mentionned in the previous section is unaffected by the NP scenario in the $B_d$ mixing assumed here).\\
When removing BR$(B^+\to\tau^+\nu)$ observable from the fit, the deviation reduces down to 0.9~$\sigma$ for both the 1D null NP-phase hypothesis and the 2D SM hypothesis.

\subsection{The $B_s$ case}

The fit for New Physics in the $B_s$ mixing also exhibits a 2.1~$\sigma$ deviation 
for the SM 2D hypothesis, $\Delta_s^{\textrm{\tiny NP}}=1$ and a 2.5~$\sigma$ for the 1D null NP-phase hypothesis, $\Phi^{\textrm{\tiny NP}}_s$=0~\footnote{A deviation in excess of 3~$\sigma$ in the fit for NP in the $B_s$ mixing had  previously been reported by the UTFit collaboration~\cite{UTfit} and has recently been updated to a lower value~\cite{Maurizio}}.
Contrarily to the $B_d$ case, the source of the discrepancy is obvious: the deviation is fully dominated by the direct measurement by the CDF and the D0 experiments of the correlated weak mixing phase and lifetime difference, ($\Delta\Gamma_s,2\beta_s$) through the time-dependent angular analysis of the $B_s\to J/\Psi\Phi$ decay . Both collaborations report a large phase with respect to the SM expectation~\cite{Phis}. The Heavy Flavor Averaging Group~\cite{HFAG} obtains a 2.2~$\sigma$ deviation for the SM hypothesis in the D0/CDF ($\Delta\Gamma_s,2\beta_s$) combination. 
The other contributions to the NP fit only provide weak constraint:
\begin{itemize}
\item the oscillation frequency, $\Delta m_s$, only sensitive to the module of $\Delta^{\textrm{\tiny NP}}_s$ is 
consistent with its SM prediction.
\item the semileptonic asymmetries, $A_{SL}^s$, and the  Flavor-Specific proper-time, $\tau_{FS}^s$ do not display any significant sensitivity to the NP fit parameters in the current state of their determination.
\item the theoretical lifetime difference proportional to the cosine of the weak phase:  $\Delta\Gamma_s \simeq 2 |\Gamma_{12}^s|\times \textrm{cos}(\Phi_s^{\textrm{\tiny NP}})$ slightly tends to push the NP phase toward the zero value as the measured $\Delta\Gamma_s$  is larger than the SM expectation for $2 |\Gamma_{12}^s|$ ~\cite{Lenz}.
\end{itemize}

\section{SUMMARY AND CONCLUSIONS}
The current data  successfully fit the KM mechanism establishing it as the dominant source of CP violation in the $B_d$ sector. The current limitation of the CKM consistency tests are of three origins: LQCD quantities, the UT angle $\gamma$ and the $B_s$ measurements.  With the recent update of the inputs,  a slight tension appears in  the global CKM fit within the SM hypothesis. This tension is mainly driven by the  larger BR$(B^+\to\tau^+\nu)$ updated value and the  smaller sin$(2\beta)$ value that B-factories recently reported and results from a non-trivial correlation that we have explicited in the ratio BR$(B^+\to\tau\nu)$/$\Delta_{m_d}$.
Both updates are consistent with the previous measurements but slightly go away from their SM prediction.
Besides the possible statistical fluctuation of the experimental measurements, further investigations are necessary to understand the role LQCD quantities may play in the origin of the tension.\\

Assuming the presence of New Physics in the $B$ mixing, a model-independent fit procedure has been applied. 
As a translation of the observed tension in the SM  CKM fit, the SM hypothesis is 2.1~$\sigma$ away from the preferred NP fit solution in the $B_d$ case.
A 2.1~$\sigma$ deviation to SM is also observed in the $B_s$ sector, fully dominated by the direct experimental measurement of the $B_s$ mixing phase by the Tevatron experiments.\\
Within a joint fit for NP in the $B_d$ and $B_s$ mixing, the 4-dimension SM hypothesis ($\Delta^{\textrm{\tiny NP}}_d = \Delta^{\textrm{\tiny NP}}_s = 1$) results in a 2.9$\sigma$ deviation with respect to the prefered fit solution.\\

A more complete update with all the data eventually available at the end of the summer 2008 is to appear in~\cite{CKMfitter}.
In the near future, significant improvements in the lattice QCD predictions as well as more precise experimental measurements can be expected in both the $B_d$ and the $B_s$ sectors. These data will be scrutinized accurately to confirm or not the appearing tension in the $B_d$ sector and the large mixing phase observed in the $B_s$ mixing. Eventually a major step in the  consistency tests of the KM mechanism will arise with the imminent LHC era and in particular with the dedicated LHCb experiment.

\begin{acknowledgments}
My warmest thanks go to my collaborators of the CKMfitter group for their support in the preparation of this talk. I also would like to thank the organizers of the ICHEP08 conference for their kind hospitality in Philadelphia, PA.
\end{acknowledgments}



\end{document}